# Discovery of charge order and corresponding edge state in kagome magnet FeGe

**Authors:** Jia-Xin Yin[1]*†, Yu-Xiao Jiang[1]*, Xiaokun Teng[2]*, Md. Shafayat Hossain[1], Sougata Mardanya[3], Tay-Rong Chang[3], Zijin Ye[4], Gang Xu[4], M. Michael Denner[5], Titus Neupert[5], Benjamin Lienhard[6], Han-Bin Deng[7], Chandan Setty[2], Qimiao Si[2], Guoqing Chang[8], Zurab Guguchia[9], Bin Gao[2], Nana Shumiya[1], Qi Zhang[1], Tyler A. Cochran[1], Daniel Multer[1], Ming Yi[2], Pengcheng Dai[2], M. Zahid Hasan[1,10,11,12]†

**Affiliations:**

[1]Laboratory for Topological Quantum Matter and Advanced Spectroscopy (B7), Department of Physics, Princeton University, Princeton, New Jersey 08544, USA.

[2]Department of Physics and Astronomy, Rice Center for Quantum Materials, Rice University, Houston, Texas 77005, USA.

[3]Department of Physics, National Cheng Kung University, Tainan 70101, Taiwan.

[4]Wuhan National High Magnetic Field Center & School of Physics, Huazhong University of Science and Technology, Wuhan 430074, China.

[5]Department of Physics, University of Zurich, Winterthurerstrasse 190, 8057 Zurich, Switzerland.

[6]Department of Electrical Engineering and Computer Science, Massachusetts Institute of Technology, Cambridge, MA 02139, USA.

[7]Beijing National Laboratory for Condensed Matter Physics and Institute of Physics, Chinese Academy of Sciences, Beijing 100190, China.

[8]Division of Physics and Applied Physics, School of Physical and Mathematical Sciences, Nanyang Technological University, Singapore 639798, Singapore.

[9]Laboratory for Muon Spin Spectroscopy, Paul Scherrer Institute, CH-5232 Villigen PSI, Switzerland.

[10]Lawrence Berkeley National Laboratory, Berkeley, California 94720, USA.

[11]Princeton Institute for the Science and Technology of Materials, Princeton University, Princeton, NJ 08544, USA

[12]Quantum Science Center, Oak Ridge, TN 37830, USA

†Corresponding authors, E-mail:

 jiaxiny@princeton.edu; mzhasan@princeton.edu
*These authors contributed equally to this work.

**Kagome materials often host exotic quantum phases, including spin liquids, Chern gap, charge density wave, and superconductivity. Existing scanning microscopy studies of the kagome charge order have been limited to non-kagome surface layers. Here we tunnel into the kagome lattice of FeGe to uncover features of the charge order. Our spectroscopic imaging identifies a 2×2 charge order in the magnetic kagome lattice, resembling that discovered in kagome superconductors. Spin-mapping across steps of unit-cell-height demonstrates the existence of spin-polarized electrons with an antiferromagnetic stacking order. We further uncover the correlation between antiferromagnetism and charge order anisotropy, highlighting the unusual magnetic coupling of the charge order. Finally, we detect a pronounced edge state within the charge order energy gap, which is robust against the irregular shape fluctuations of the kagome lattice edges. We discuss our results with the theoretically considered topological features of the kagome charge order including unconventional magnetism and bulk-boundary correspondence.**

A kagome lattice, made of corner-sharing triangles, is a geometrically frustrated lattice first introduced to quantum physics [1] in 1951, while a similar pattern unit has long been used as the star of David in religious ceremony and as the Hexagram in alchemy symbols. With the recent advance of quantum materials, it has been demonstrated that kagome materials can support a variety of quantum phases, including spin liquids, Chern gap, charge density wave, and superconductivity, significantly promoting the diversity of quantum matter in a model lattice setting [1-6]. In contrast to other low-temperature phases, the kagome charge order can occur at high temperatures up to 100K exhibiting many-body correlations and Berry curvature physics. Recent theories [6-13] have proposed topological features of the kagome charge order, including orbital currents/magnetism, and bulk-boundary correspondence. Fundamentally, the charge order driven by the kagome van Hove singularities can induce a nontrivial Berry phase and accordingly various Berry phase related quantum effects [14], presenting a rare case of interaction-driven quantum topology explored in recent theories [15,16,17]. The many-body interaction not only opens a charge density wave gap at $E_F$, but also produces a nontrivial Berry phase that leads to orbital currents/magnetism and a topological edge state, as illustrated in Fig. 1(a) [7-13].

Scanning tunneling microscopy (STM) was among the first experimental techniques to discover the charge order in kagome superconductors $AV_3Sb_5$ (A=K, Rb, Cs). However, existing STM studies have been focused on the Sb honeycomb layers [7,18-26], rather than directly probing the kagome lattice layer. Notably, intriguing kagome physics may not be detected on the non-kagome surface layers [4,27,28], and features detected on non-kagome layers (such as surface stripe reconstructions) may also not be directly related to the kagome lattice. Therefore, tunneling directly into the kagome lattice is essential to probe the intrinsic features of the kagome charge order, and is particularly important to solving puzzles regarding different behaviors of the charge order on non-kagome surfaces [7, 18-26]. Recently, our combined neutron scattering [29], STM and photoemssion experiments on kagome antiferromagnet FeGe unexpectedly detected a superlattice signal around 100K, suggesting it likely to be a kagome charge order platform. According to the STM study on CoSn — a cousin material to FeGe — the atomic kagome lattice layer can be prepared as the surface termination through cryogenic cleaving [28], offering us an unprecedented opportunity to explore the kagome charge order at the microscopic level.

In this work, single crystals with a size up to 2mm × 2mm are cleaved in situ at 77K in ultra-high vacuum conditions and then inserted into the microscope head, already at $^4$He base temperature (4.2K). Data are taken at 4.2K unless otherwise specified. Topographic images in this work are taken with the tunneling junction set-up $V = 60$mV and $I = 0.05$nA. Tunneling conductance spectra are obtained with an Ir/Pt tip using standard lock-in amplifier techniques with a lock-in frequency of 977Hz and a junction set-up of $V = 100 \sim 60$mV, $I = 1$nA, and a root mean square oscillation voltage of 0.5mV. Tunneling conductance maps are obtained with taking a spectrum at every location under a junction set-up of $V = 60$mV, $I = 0.3 \sim 0.1$nA, and a root mean square oscillation voltage of 5mV.

FeGe has a hexagonal structure (space group P6/mmm) with lattice constants $a = 5.0$ Å and $c = 4.0$ Å. It consists of a $Fe_3Ge$ kagome layer and a $Ge_2$ honeycomb layer with alternating stacking [Fig. 1(b)]. Pioneering studies [30] have established it as a kagome antiferromagnet with a Neel transition over 400K. The spins of Fe kagome lattice are ordered predominantly along the *c*-axis, while a weak spin canting has been observed at low temperatures [30]. The cryogenic cleaving of 11 systems produces both the honeycomb lattice and kagome lattice layers as the natural terminations [Fig. 1(b)]. Figure 1(c) shows a case of a half-unit-cell step with a step height of 2.0Å, revealing both $Ge_2$ honeycomb layer and $Fe_3Ge$ kagome layer. In this work, we focus on the kagome lattice, as shown in Fig. 1(d), and its inset. Similar to the cases in CoSn [28], and $Fe_3Sn_2$ [27,31], the individual atoms in the honeycomb layer can be well resolved, while the individual kagome atoms are often not easy to be resolved in the STM experiment. We further perform spectroscopic imaging on the kagome lattice, focusing on the Fermi-level electronic structure. The Fourier transform of the dI/dV map in Fig. 1(e) and its inset demonstrates the 2×2 vector peaks from the charge order in addition to the Bragg peaks of the lattice. The 2×2 vector peaks are consistent with the wave vector of bulk charge order obtained in our neutron study of FeGe. The three pairs of 2×2 vector peaks feature different intensities as shown in Fig. 1(f), resembling the chiral charge order reported in the defect-free Sb surfaces in $AV_3Sb_5$. Counting from the highest to the lowest intensity vector peak pairs, we define an anticlockwise chirality [7,19,20,24] for this atomic region. At $E_F$, we also detected an energy gap with a size around 50meV [Fig. 1(g)]. Across this energy gap, a partial real-space charge density reversal is observed through spectroscopic imaging [Fig. 1(h) inset], as expected for a charge density wave order [7,34]. This gap disappears above the charge ordering temperature that is around 100K, together with the disappearance of the 2×2 vector peaks in the spectroscopic map data, as shown in Fig. 1(h) and its inset. These spectroscopic data sets establish the existence of charge order in the magnetic kagome lattice of FeGe.

An outstanding question is whether our observed charge order can be topological as proposed in several related theories [6-13]. STM can be used to examine the nontrivial consequences of topological phases. Theoretical treatment of chiral charge order often considers a winding phase (tied to its chirality) between the three sets of order parameters associated with three symmetry-related ordering vectors. This phase difference of the triple order parameters leads to orbital currents with a nontrivial Berry phase [Fig. 1(a)][6-13]. The associated Berry curvature field introduces a weak orbital magnetization with its direction tied to the chirality. Meanwhile, the proposed topological charge order can feature a bulk-boundary correspondence with edge states inside the bulk gap. Specifically, in the case of the antiferromagnet FeGe, one can expect that the weak orbital magnetization of the topological charge order can be polarized by the strong antiferromagnetic spin order, resulting in a chirality-switching charge order between adjacent kagome layers that our tunneling experiments can directly reveal.

To explore the interplay between charge order and dominant spin order in the kagome lattice, we perform spin-polarized tunneling. While spin-polarized tunneling signals have been detected in kagome ferromagnet [35,36], spin-polarized tunneling measurements are lacking for kagome antiferromagnets. One of the three ways for preparing magnetic probes is to dip the nonmagnetic tip into magnetic materials [37]. We find that progressively dipping the Pt/Ir tip into the magnetic kagome layer can effectively produce a magnetic tip. With this magnetic tip, we study a nearby region composed of kagome surfaces with unit-cell-height steps [Fig. 2(a)]. The spin-flip field for FeGe is known [29] to exceed 10T, beyond our instrumental capability, and thus we can safely determine that our low magnetic field only flips the spin of the magnetic tip. Figure 2(b) shows dI/dV maps taken at $E_F$ for the same nano-region in Fig. 2(a) with the tip spin polarized in the up direction (upper panel) and down direction (down panel), respectively. In the upper panel, terraces ②④⑥ have a stronger intensity, while in the lower panel, terraces ①③⑤⑦ reveal a more prominent intensity. The intensity contrast and contrast reversal systematically provide key evidence for spin-polarized tunneling. Figure 2(c) compares the dI/dV signals for terraces ④ and ⑤ as sweeping $c$-axis magnetic field from -1T to +1T. A switch of their signals for positive and negative fields is obtained, and further indicates that the saturation field of the magnetic tip is around 0.4T. Our data sets suggest that the kagome lattice has strong spin-polarized states at $E_F$ and the adjacent kagome layers have states with opposite dominant spin polarizations, which are fully consistent with the expected bulk antiferromagnetic spin order along the $c$-axis.

With the demonstration of antiferromagnetism of FeGe and the spin-resolving capability of our magnetic probe, we further characterize the magnetic features of the charge order. We zoom into the boxed area for terraces ④ and ⑤ and measure additional atomic-scale dI/dV maps with the tip spin polarized to opposite directions. We subtract the two maps taken with opposite tip polarizations to obtain the magnetic contrast map in Figs. 2(d) and (f) after carefully aligning their atomic positions. Both magnetic contrast maps show 2×2 modulations, as further evidenced by their Fourier transform in Figs. 2(e) and (g). While both magnetic contrast maps show 2×2 charge order, we find their chirality is evidently opposite to each other [Figs. 2(e) and (g)], thus defining an antichiral charge order. Theoretical proposals [6-13] of topological charge order often indicate (Berry curvature induced) weak orbital magnetism [38-40] whose direction is tied to the chirality of the underlying order parameter and the orbital magnetization moment can couple to either an external or internal magnetic field. In $AV_3Sb_5$, the applied magnetic field can switch the chirality of the charge order. In contrast, in FeGe, the strong staggered internal magnetic field can cause switching of the chirality between different kagome layers. Such a magnetic coupling of the charge order is consistent with our neutron scattering observation [29] of additional enhancement of the magnetic moment just below the charge ordering temperature. In addition, the chirality switch of the charge order is similarly evidenced by using a nonmagnetic tip to image a unit cell step [Fig. 2(h)], further confirming our spin-polarized tunneling results.

Besides the orbital magnetism, theoretical studies [6-13] of the charge order also suggest the existence of boundary modes resulting from a topological charge order bulk gap. The bulk-boundary correspondence is a key concept in quantum topology [41]. In a topological material, the nontrivial bulk gap features gapless boundary states. Through extensive scanning searching, we can obtain unit-cell-height kagome step edges, as shown in Fig. 3(a). Spectroscopic imaging at $E_F$ with a nonmagnetic tip at zero-field clearly demonstrates pronounced edge states for all these edges. The edge state is rather robust in that it is observed almost independently of the irregular shape of the edges, pointing to a protection mechanism. We do not detect edge states for energies outside the charge order gap, as demonstrated in the series of dI/dV mapping data in Fig. 3(b). The high-resolution dI/dV spectrum taken at the edge in Fig. 3(c) confirms the emergence of states only inside the charge order gap. We further show the spatial decay of the edge state towards the kagome lattice in Fig. 3(d). An exponential fitting of the decay curve estimates the characteristic decay length of 2.1nm. We also confirm that the edge state disappears above the charge ordering temperature, as shown in Fig. 3(e). Crucially, the emergence of robust edge states within the kagome lattice charge order gap is consistent with the bulk-boundary correspondence expected for a topological charge order proposed in several theoretical works [6-13]. Compared with the topological edge states [4,42-46] seen in other metallic systems, the edge states in FeGe uniquely emerge at the $E_F$. The emergence of Fermi-level edge states implies the charge-order-driven anomalous transverse transports, which is consistent with our observation of large anomalous Hall effect just below the charge ordering temperature [29] when giant magnetic fields are applied to overcome antiferromagnetism along the *c*-axis. Taken together with the magnetic coupling of the charge order supported by the spin-chirality correlation within each kagome layer, our tunneling work provides microscopic experimental data consistent with the orbital magnetism and bulk-boundary correspondence for the charge order in kagome magnet FeGe.

**Acknowledgement:**

M.Z.H. acknowledges support from the US Department of Energy, Office of Science, National Quantum Information Science Research Centers, Quantum Science Center and Princeton University. M.Z.H. acknowledges visiting scientist support at Berkeley Lab (Lawrence Berkeley National Laboratory) during the early phases of this work. Theoretical and STM works at Princeton University was supported by the Gordon and Betty Moore Foundation (GBMF9461; GBMF4547; M.Z.H.). The theoretical work including ARPES were supported by the US DOE under the Basic Energy Sciences program (grant number DOE/BES DE-FG-02-05ER46200; M.Z.H.). T.-R.C. was supported by the Young Scholar Fellowship Program under a MOST grant for the Columbus Program, MOST111-2636-M-006-014, the Higher Education Sprout Project, Ministry of Education to the Headquarters of University Advancement at the National Cheng Kung University (NCKU), the National Center for Theoretical Sciences (Taiwan). The work at Rice was supported by US NSF-DMR-2100741, the Robert A. Welch Foundation under grant no. C-1839, C-2024, and No. C-1411, the U.S. Department Of Energy (DOE) grant no. DE-SC0021421 and No. DE-SC0018197, and the Gordon and Betty Moore Foundation's EPiQS Initiative through grant no. GBMF9470. The work at Nanyang Technological University was supported by the National Research Foundation, Singa- pore under its Fellowship Award (NRF-NRFF13-2021-0010). T.N. acknowledges support from the European Union's Horizon 2020 research and innovation programme (ERC-StG-Neupert-757867-PARATOP). G. X. acknowledges support from the National Key Research and Development Program of China (2018YFA0307000), and the National Natural Science Foun- dation of China (11874022).

## Figures

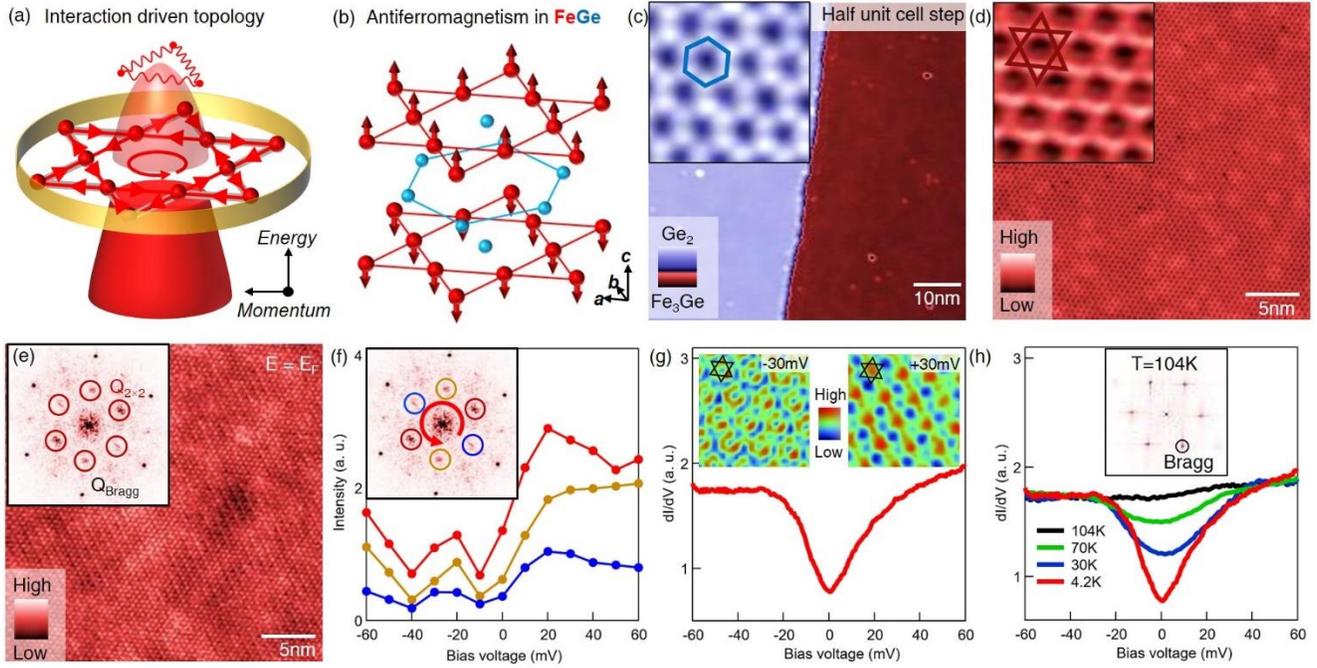

Fig 1. (a) Many-body interaction (red springs) driven topology in kagome lattice (red spheres). The interaction not only opens a charge density wave gap at $E_F$, but also produces a nontrivial Berry phase (red circular arrows) that leads to orbital currents/magnetism (thick red arrows) and a topological edge state (yellow ring). The light and dark red ellipsoid represent unoccupied and occupied electronic states, respectively. (b) Crystal structure of FeGe. The low temperature magnetic structure is an approximate A-type antiferromagnet. (c) Topographic image of a half unit cell step consisting of Ge and FeGe layers, and the zoom-in Ge layer marked with the honeycomb lattice (blue lines). (d) Topographic image of the kagome lattice layer of FeGe, and the zoomed-in image marked with the kagome lattice (red lines). (e) dI/dV map taken at $E_F$ for the same kagome lattice region, and its Fourier transform identifying the 2×2 vector peaks. (f) Energy distribution of the three pairs of 2×2 vector peak intensities. Counting from the highest to the lowest intensity vector peak pairs, we define an anticlockwise chirality. (g) dI/dV spectrum taken on the kagome lattice, showing an energy gap. The insets show two tunneling current maps for the same region taken below and above this energy gap. We also mark the kagome lattice as black lines. The center of this marked kagome lattice shows a reversal of intensity for these two maps, consistent with a charge order. (h) Temperature dependence of tunneling spectrum showing the disappearance of energy gap at 104K, which is just above the charge ordering temperature. The inset shows the Fourier transform data of a dI/dV map taken at $E_F$ at 104K, showing the disappearance of 2×2 vector peaks.

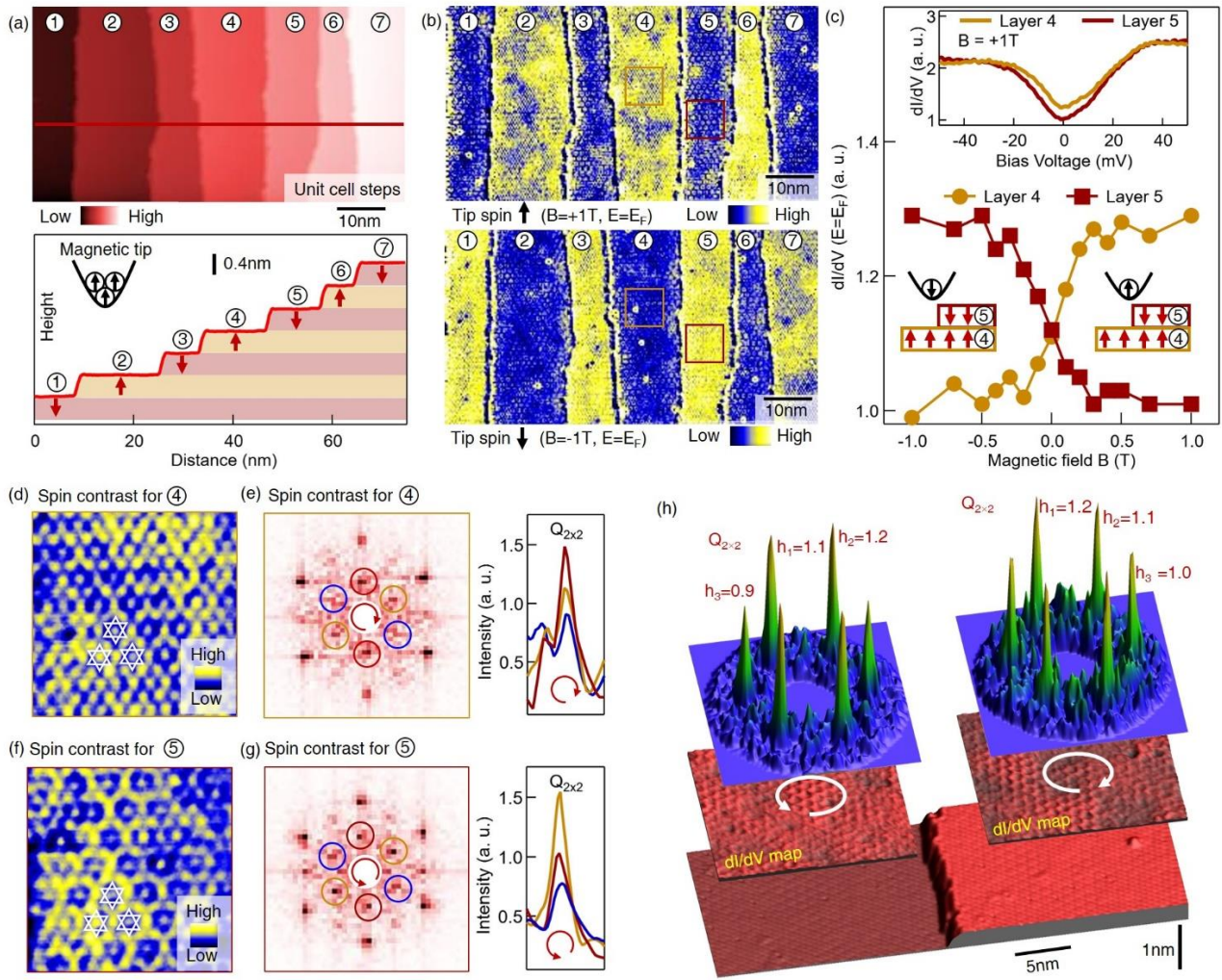

Fig 2. (a) Topographic image of kagome terraces with unit cell steps (upper panel) and their height profile (lower panel). A magnetic tip is used to obtain the data, and the kagome lattices are expected to couple antiferromagnetically along the c-axis. (b) dI/dV maps taken at $E_F$ for the region in **a** with tip spin polarized in up (upper panel) and down (lower panel) directions. (c) dI/dV signal at $E_F$ for kagome terrace ④ and ⑤ as a function of c-axis magnetic field sweeping. The lower inset shows the schetch illustrating spin configurations for the tip and sample at different magnetic fields. The upper inset shows the dI/dV data for kagome terrace ④ and ⑤ with B=+1T. (d) High-resolution dI/dV magnetic contrast data for terrace ④. The data is taken at the orange box region in (b), and obtained through subtracting +1T map with -1T map data. The white lines mark the 2×2 charge order. (e) Fourier transform of (d) showing 2×2 vector peaks. The right panel displays the intensity of the 2×2 vector peaks along the Bragg spot direction, from which we determine the clockwise chirality. (f) High-resolution dI/dV magnetic contrast data for terrace ⑤. The data is taken at the red box region in (b), and obtained through subtracting -1T map with +1T map data. (g) Fourier transform of (f) showing 2×2 vector peaks. The right panel displays the intensity of the 2×2 vector peaks along Bragg spot direction, from which we determine anticlockwise chirality. (h) Confirmation of the chirality switch with a nonmagnetic tip at zero field. The lower layer shows kagome lattices with a unit cell step. The middle layer shows dI/dV maps taken at $E_F$ for the two terraces. The upper layer shows

the 2×2 vector peaks obtained by Fourier transforming the respective map data, from which we determine the opposite chirality for the two terraces.

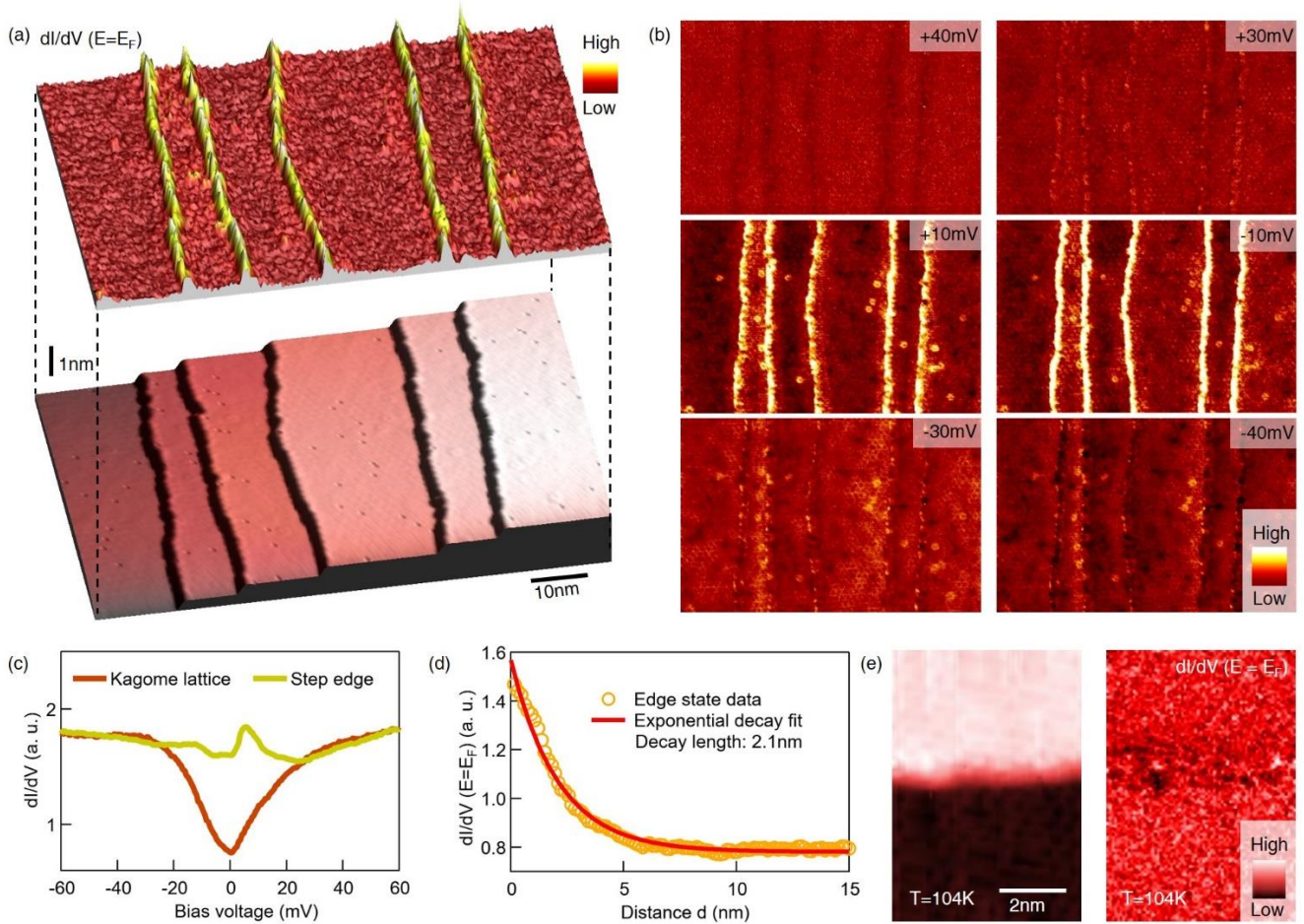

Fig 3. (a) dI/dV map taken at $E_F$ (upper layer) for unit cell kagome step edges (lower layer), showing the robust edge states. (b) dI/dV maps taken for the same region at energies outside (upper and lower panels) and within (middle panels) the charge order energy gap. The kagome edge state appears only at the energies within the charge order energy gap. (c) dI/dV spectra taken at the kagome lattice and step edge, respectively. The step edge data shows the emergence of states within the charge order energy gap. (d) Spatial decay of the edge state. The decay can be fitted with an exponential function with a characteristic decay length of 2.1nm. (e) Topographic image of a unit cell kagome step edge (left) and its corresponding dI/dV map at Fermi energy taken at 104K, showing the disappearance of edge state above the charge ordering temperature that is around 100K.

**SUPPLEMENTARY**

We organize the supplementary materials in a question and answer way, where the general comments and questions are contributed by the reviewers and editors of this work.

In this manuscript, Yin et al. have investigated the antiferromagnetic order and a derived charge order in the kagome layer of FeGe. A 2×2 charge order has been discovered, resembling that discovered in kagome superconductors $AV_3Sb_5$, and spin-mapping across steps of unit-cell-height demonstrates that it emerges from spin-polarized electrons with an antiferromagnetic stacking order. The chirality of charge order anisotropy was correlated with the antiferromagnetism, indicating the unusual magnetic coupling of the charge order. Moreover, a robust pronounced edge state appears within the charge order energy gap, which was assigned as the bulk-boundary correspondence in a topological system. These measurements are of high quality and convincing, and the conclusions given by the authors is basically consistent with their observations, offering new understanding of the interplay of CDW and antiferromagnetism in FeGe. Study of correlation driven topological phenomena in kagome lattice is a hot topic, and the presented findings and experiments are in my opinion of interest for the broader community.

Yin et al. propose a scanning tunneling microscopy and spectroscopy study of a kagome lattice. Combining usual STM and spin-polarized STM, they observe concomitant charge order and AF interlayer coupling in in-situ cleaved FeGe. Low temperature cleaving allowed the authors to expose the $Fe_3Ge$ kagome layer to the surface rather than the $Ge_2$ honeycomb layer. This enables them to directly probe the electronic properties of the kagome lattice which are not accessible when tunneling through a $Ge_2$ surface layer. The key results of this study are a 2x2 charge density wave reconstruction in the $Fe_3Ge$ kagome layer with a gap of about 50meV. Each layer is ferromagnetic with an antiferromagnetic (AF) coupling between adjacent layers. The AF layer stacking is inferred form spin-polarized STM across a step edge for different tip polarizations (tuned via an external applied magnetic field between -1T and 1T). perpendicular to the kagome surface. The authors further find a robust edge state at the fermi level associated with the charge order gap. Finally, they claim that the Charge order and magnetic order arechiral, based on the analysis of the 2x2 lattice peaks in the Fourier transform of conductance maps taken with a normal tip and a spin polarized tip.

Technical questions & answers:

**Question 1: In kagome superconductor $KV_3Sb_5$, a recent STM study [Nature Physics 18, 265–270 (2022)] shows nematicity of the CDW and attributes the chiral CDW to tip effect; how do the authors respond?**
Answer: We think the discussion of this topic is indeed relevant for the understanding of chirality and nematicity in FeGe.

First, we broadly comments on the topic of nematicity in the kagome superconductor $AV_3Sb_5$ in light of original publications from Physical Review and other journals.
1) The 2×2×2 CDW in $KV_3Sb_5$ was initially established by several works. Both magnetization and specific heat measurements of $KV_3Sb_5$ find an anomaly at T*, which is initially discussed as "The possibility of an orbital-ordering event is considered" [quoted from abstract of Rev. Mater. 3, 094407 (2019)]. With discovery of low-temperature superconductivity of $CsV_3Sb_5$, an in-plane superlattice structure is also detected below T* through X-ray scattering, which is suggested to be due to a "charge density wavelike instability" at the van Hove band filling ["charge density wavelike instability" quoted from Phys. Rev. Lett. 125, 247002 (2020)]. Scanning tunneling microscopy [Nat. Mater. 20, 1353–1357 (2021)] identifies the 2 × 2 in-plane wave vector of the charge modulation, charge energy gap, and in-plane 2 × 2 charge density reversal across the gap of the T* phase in $KV_3Sb_5$, which reveals a charge density wave order. X-ray scattering in supplementary of Nature Materals (2021) further identifies the 2 × 2× 2 three-dimensional wave vectors of the T* phase, following the X-ray work for $RbV_3Sb_5$ [Phys. Rev. X 11, 031050 (2021)]. These systematic experiments taken together establish the bulk 2 × 2 × 2 charge density wave as the many-body state of T* phase $KV_3Sb_5$, and the $2 \times 2$ charge density wave vector agrees with the initial considerations of electronic instability for kagome lattices [Phys. Rev. B 87, 115135 (2013); Phys. Rev. Lett. 110, 126405 (2013)].

2) For a bulk CDW order with a similar superlattice of $2 \times 2 \times 2$, nematicity has been reported in $TiSe_2$ by Phys. Rev. Lett. 105, 176401 (2010) entitled "Chiral charge-density wave": "The chirality in CDW results in the threefold symmetry breaking. Macroscopically, twofold symmetry was indeed observed in optical measurement." as we quoted from the PRL 2010. The existence of chiral CDW in $TiSe_2$ has been reconfirmed by the bulk chiral optical manipulation of the CDW order [Nature 578, 545–549 (2020)]. However, we still don't know whether the chiral CDW in $TiSe_2$ breaks time-reversal symmetry or not.

3) Before the discovery of $AV_3Sb_5$ [Rev. Mater. 3, 094407 (2019)], nematicity in kagome lattice has been reported in kagome magnet $Fe_3Sn_2$ by STM and magnetotransport [Nature 562, 91–95 (2018), Phys. Rev. Lett. 123, 196604 (2019)]: "Probing the fermionic quasi-particle interference reveals consistent spontaneous nematicity—a clear indication of electron correlation," as quoted from the abstract of Nature 2018.

4) The nematicity in $AV_3Sb_5$ is initially reported by magnetotransport Chin. Phys. Lett. 38, 057403 (2021), where the measuring geometry is similar to PRL 2019 for $Fe_3Sn_2$. Another magneto-transport confirms the nematicity [Nat Commun 12, 6727 (2021)] with its measuring geometry similar to the magnetotransport in the extended data of Nature 2018 for $Fe_3Sn_2$.

5) Prior to the publication of Nature Phys 2022, both chirality and nematicity of the CDW order have been reported in $AV_3Sb_5$ by STM: Supplementary in Nat. Mater. 20, 1353–1357 (2021), main figure in Phys. Rev. B 104, 035131 (2021), main figure in Phys. Rev. B 104, 075148 (2021). In these three STM papers, surface regions featuring chirality and nematicity are both shown for all three $AV_3Sb_5$ (A=K, Rb, Cs) compounds, and the absence of chirality is found to be related to pinning by defects. Actually, a smoothing/blurry of the chirality data will result in nematicity (lowering q-resolution), as noted by Phys. Rev. B 104, 035142 (2021): "To our eyes, their Fourier transform STM data (Fig.3abc of Nature Materials 2021) is better described by two strong and approximately equal intensities and one weaker one, which is consistent with our theory of inter-layer coupling".

Secondly, we note that the STM observation of chiral CDW with magnetic field tunable chirality in $AV_3Sb_5$ is able to make connections to several bulk sensitive experiments beyond STM's own capability.

1) These experiments including: the anomalous Hall transport effect [Sci. Adv. 6, eabb6003 (2020), Physical Review B 104, L041103 (2021)], time-reversal symmetry breaking of the CDW by muon spin rotation and Kerr rotation [Nature 602, 245-250 (2022), arXiv:2107.10714 (2021), arXiv:2110.11306 (2021), arXiv:2202.07713 (2022), Phys. Rev. Research 4, 023244 (2022), arXiv: 2204.10116 (2022, to appear in Nature Physics)], chirally anisotropic CDW vector peaks by hard X-ray scattering [Phys. Rev. B 105, 195136 (2022)], field-tuned nonlinear transport [arXiv:2203.09593 (2022 to appear in Nature)].

2) The link between STM detection of magnetic field tunable chiral CDW and these bulk experiments is substantially discussed by several theoretical work [Sci. Bull. 66, 1384–1388 (2021); Phys. Rev. Lett. 127, 217601 (2021); Phys. Rev. B 104, 045122 (2021); arXiv:2105.15204 (2021); arXiv:2110.06266 (2021); Preprint at arXiv:2109.10809 (2021); Nat. Phys. 18, 137–143 (2022)]. These theoretical analysis connect to original kagome charge order instability analysis in Phys. Rev. Lett. 110, 126405 (2013), as well as orbital currents proposal in other lattices including Haldane model for producing quantum anomalous Hall effect [Phys. Rev. Lett. 61, 2015–2018 (1988) and Varma model for loop currents in cuprates [Physical Review B 55, 14554-14580 (1997)].

3) The STM results reporting the absence of chirality cannot make a reasonable association with these bulk experiments reporting the chiral and magnetic nature of the CDW. By contrast, chiral CDW can be naturally consistent with the bulk experiments claiming nematicity [see original work as an example: Phys. Rev. Lett. 105, 176401 (2010)]

4) In recent nonlinear transport work [arXiv:2203.09593 (2022, to appear in Nature)], it is demonstrated that the intrinsic field tuned chirality is detected in strain-free devices, and strain can destroy the chiral signal.

This contrast is argued to be able to reconcile the difference between STM works in Nature materials 2021 and Nature Physics 2022.

Thirdly, we make a direct comparison between STM data in Nature Materials 2021 and Nature Physics 2022, and clarify their key differences.

1) The first clear difference between their STM data is the CDW gap. STM in Nature materials 2021 shows a large gap opened from -25meV to +25meV, which is consistent with ARPES data [Nature Comm 13, 273 (2022)] that the occupied states near M open a gap of 25meV. However, the STM in Nature Physics 2022 shows a much smaller CDW gap from -5meV to +10meV. Actually, this small gap is quite consistent with the defect-rich region (showing absence of chirality) discussed in supplementary of Nature materials 2021. A comparison of the data from the two works shows remarkable similarity (as compared below). In reference to ARPES data, this smaller gap suggests tunneling more to the Sb p orbital states (central round circular Fermi surface) rather than the V d orbital states.

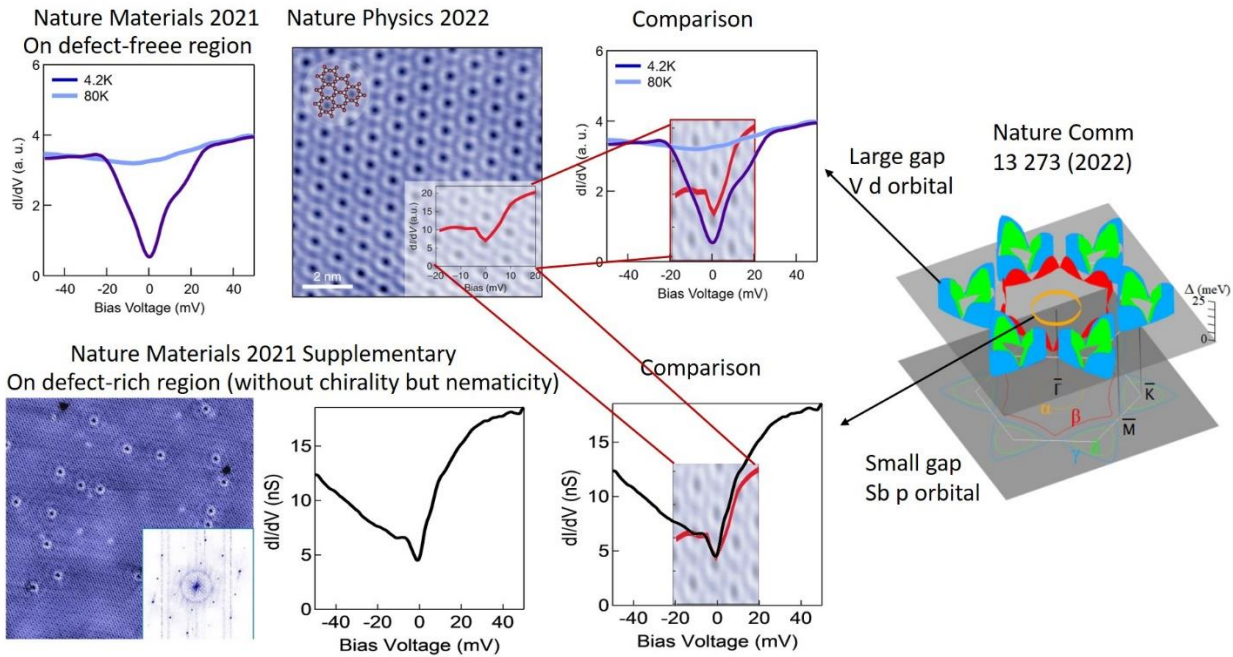

2) In the supplementary of Nature Materials 2021, it has been shown that in the defect-rich region, there is no chirality but nematicity owing to impurity pinning of the CDW. The impurity pinning phase of CDW has been discussed in classic work by Lee, Rice, Anderson, Fukuyama et al. for 1D CDW system [Solid State Communications 14, 703 (1974); Physical Review B 17, 535 (1978); Physical Review B 19, 3970 (1979)], and a direct STM demonstration of these theories is reported recently [Phys. Rev. Lett. 126, 256401 (2021)]. Note that in Nature materials 2021, the defect-rich region also contains a small clean region ~5nm×5nm that is surrounded by defects. While the defect-free region in Nature Materials 2021 refers to regions without impurities introducing standing waves for clean areas larger than 30nm×30nm. In Nature Physics 2022, the so-called "clean region" is shown to be ~5nm×5nm. In the Fourier transform of the so-called "clean region", there are still detectable ring-like signals, which are known to arise from defect/disorder-induced scattering (as marked below). The existence of a ring-like signal in the Fourier transform data in Nature Physics 2022 is in sharp contrast with the data showing chirality in Nature materials 2021. The data presented in Nature Physics 2021 again resembles the "dirty" region as discussed in the supplementary of Nature Materials 2022. The ring-like signal is also known to arise from the intra-band scattering of Sb p orbitals with a circular Fermi surface [reconfirming our point 1), see the corresponding circular Fermi surface below], which is not directly related with V-d orbital kagome physics.

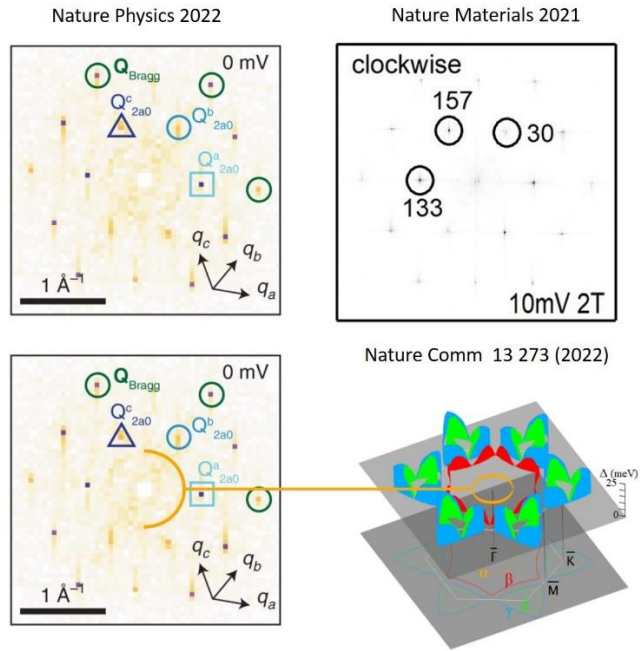

3) We examine the proposal in arXiv:2203.09593 (2022, to appear in Nature) that it is proposed that the strain in the sample can destroy chirality. We find similarly that large surface area data in Nature Physics 2022 does show strong hints to strain force, as indicated by the large surface inhomogeneity (white areas as bumps with unkown origin). For the smaller surface areas claimed to be "clean", we find the Q = 0 data is removed/filtered by their authors, hindering us from making an evaluation. For their data shown in extended data, it does show similar strong strain force effect possibilities (as indicated by similar rich bump areas on topographic data). In addition, in the extended data of Nature Physics 2022 the chirality switch is mostly examined through topographic image; while in Nat. Mater. 20, 1353–1357 (2021), Phys. Rev. B 104, 035131 (2021), Phys. Rev. B 104, 075148 (2021), the chirality and magnetic field switching are demonstrated in STM dI/dV maps instead of topographic images.

Forthly, in the extended data of Nature Physics 2022, there is a case of chirality switch on topographic data, and is attributed to tip effect. We discuss the differences between this data and other data establishing chirality.

1) In previous STM work [Nat. Mater. 20, 1353–1357 (2021), Phys. Rev. B 104, 035131 (2021), Phys. Rev. B 104, 075148 (2021)], chirality is established through analyzing the dI/dV maps (that are more sensitive to electronic order), rather than topographic images that is more sensitive to lattice effects.
2) There is no fundamental principle forbidding an STM tip to induce a chirality switch. For instance, in Nature Communications 13, 1843 (2022) entitled "Atomic-scale visualization of chiral charge density wave superlattices and their reversible switching", it is shown that STM can switch a chiral CDW. The mere observation of chirality switch by tip cannot role out the existence of intrinsic chiral CDW.
3) Prior to the publication of Nature Physics 2022, the magnetic field switch of the chirality of CDW supports the intrinsic nature rather than arising from an anisotropic tip [Nat. Mater. 20, 1353–1357 (2021), Phys. Rev. B 104, 035131 (2021)].
4) Prior to the publication of Nature Physics 2022, the chirality of CDW has been detected directly in the CDW-gap map in $RbV_3Sb_5$ and $CsV_3Sb_5$ [Phys. Rev. B 104, 075148 (2021)], which directly measures the energy modulation that is free from set-point effect and other tip anisotropy effects. The chirality in CDW gap map supports the intrinsic nature of chirality.

Lastly, despite all the reasoning listed above, we agree that all the STM discussions in $AV_3Sb_5$ system are based on non-kagome surface measurement, which all can potentially not directly related with the underlying intrinsic kagome physics. Instead, as shown in this work, STM can directly work on the kagome lattice of FeGe offering new opportunities to investigate the intrinsic nature of the CDW order, that can be further discussed beyond this work.

**Question 2**: Some data has been mentioned in the main text but was not provided, and the omitted data are including: (1) "We have also checked spectroscopic imaging of unit-cell steps with a nonmagnetic tip under a similar magnetic field, which does not show an intensity difference between adjacent terraces."(2) "We subtract the two maps taken with opposite tip polarizations to obtain the spin contrast map in Figs. 2(d) and (f) after carefully aligning their atomic positions." The two maps with opposite tip polarizations and the details of the alignment of atomic positions should be provided.

Answer:
1) We find the spectrum taken at adjacent kagome layers (noted as A&B) under 1T applied along $c$-axis is identical with each other with a nonmagnetic tip:

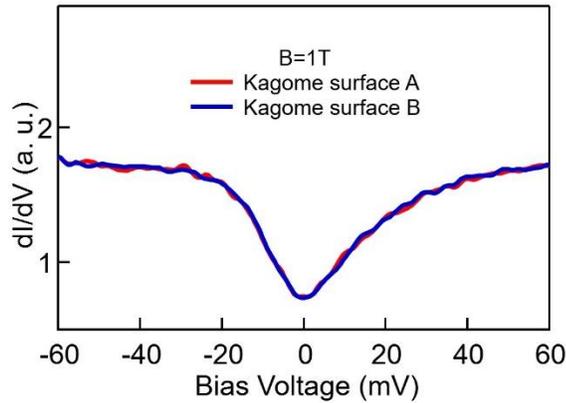

2) The two maps are identical to that presented in Fig. 2b, but are of a smaller spatial scale, which are reproduced below. The centers of two defects are used to align the atomic positions of the two maps. It is visible from the two maps that the atomic patterns are reversed for surface 4 and surface 5 through switching the magnetic field.

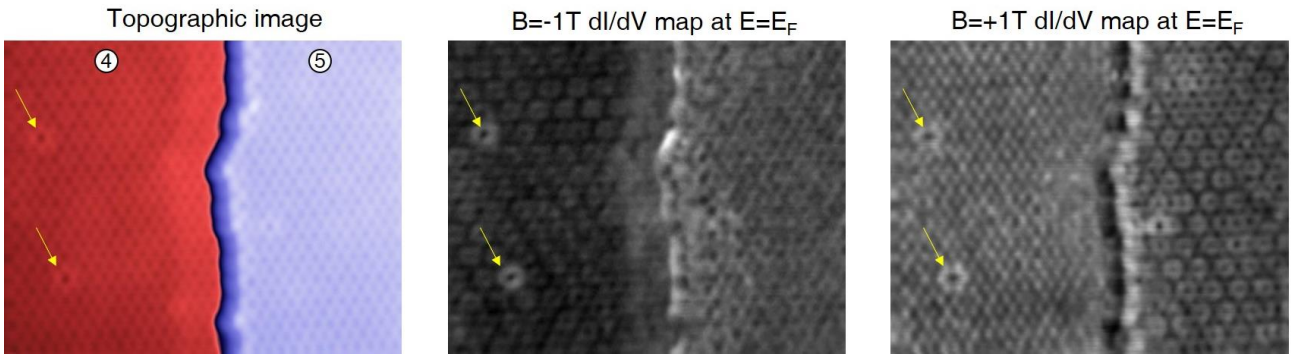

We also note that the spin-polarized map is taken at $E_F$ within the CDW gap, which makes it different from a conventional spin-polarized signal.

Suppose we have two kinds of electrons with spin-up and spin-down polarizations, while the CDW gap opens in the spin-up channel but not the spin-down channel. For spin-polarized tunneling, when we sense higher intensity at $E_F$, it indicates tunneling more to the spin-down channel, rather than the spin-up channel. If this is the case in our experiment, then it leads to the opposite spin-polarized tunneling behavior between the kagome lattice state at $E_F$.

That is to say, when the spin maps have lower intensity, the tip is more sensitive to the spin-polarized electron channel where CDW gap develops better, and the 2×2 charge order modulation is stronger, consistent with our data and interpretation.

**Question 3: The reported data were measured in 4K, where a canting AFM state exists, which contributes the opposite in-plane FM component. Do this FM components affect or induce the CDW chirality?**

Answer: In the $AV_3Sb_5$ kagome system where there is no AFM order with or without canting, the chiral CDW has been reported, spontaneously breaking time-reversal symmetry as evidenced by muon spin rotation [Nature 602, 245-250 (2022)]. Therefore, a FM component of magnetic order may not be a key reason to induce the chiral CDW. Theoretical studies of chiral CDW in kagome systems indicate the existence of a weak orbital magnetization along c-axis whose direction is coupled to the chirality of CDW [Nat. Mater. 20, 1353–1357 (2021); Sci. Bull. 66, 1384–138 (2021); Phys. Rev. Lett. 127, 217601 (2021); Phys. Rev. B 104, 045122 (2021)]. This is also regarded as the time-reversal symmetry-breaking nature of the chiral CDW.

For the in-plane FM components, we admit that we are not sure how it will interact with the chiral CDW. In principle, if there are in-plane orbital currents/magnetization induced by the stacking of chiral CDW along c-axis, the in-plane FM can couple to that chiral component of the three-dimensional CDW. A STM study based on side cleaving of the crystal may be necessary to clarify this physics.

Meanwhile, we also think it is meaningful to ask whether the chiral CDW affects the magnetic order. In pioneering neutron studies, it has been established that the in-plane FM forms an anti-spiral order along c-axis in FeGe. Its chirality, interestingly, correlates well with the antichiral CDW order reported in this work, and the CDW is established as the parent phase of the anti-spiral magnetic order. The rich intertwining between magnetism and charge order definitely deserves substantial future work, in theory and experiment.

**Question 4: Fig. 3 show pronounced edge states for all unit-cell step edges, which was considered as the bulk-boundary correspondence. However, it seems that the point defects also possess in-gap states and show similar energy-dependent behavior. How to understand? Besides, what about the edge states of half-unit-cell edges?**

Answer: We have not studied the defect effects carefully in FeGe, which would be a valuable direction. We find two differences in our existing data:

1. The impurity state is weaker than the edge state at $E_F$. We use white arrows to mark several impurity states away from edges in the image shown below:

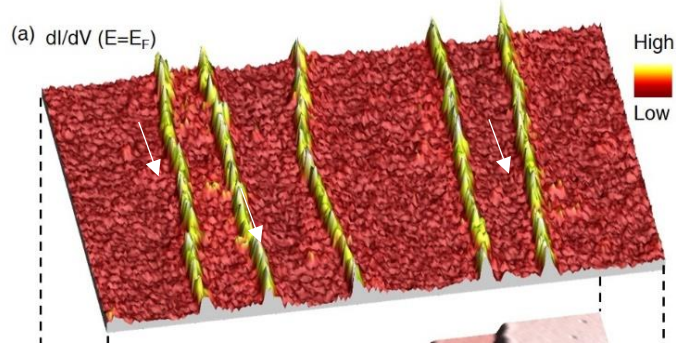

2. In the -30meV and -40meV dI/dV map data as reproduced below, some of the defect states are still very pronounced, while the edge state almost disappears.

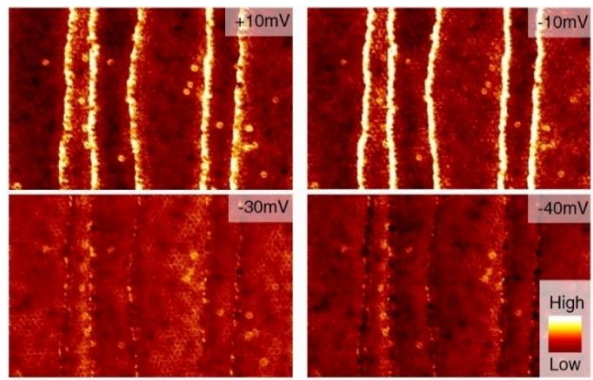

Defect in kagome lattice materials is known to introduce novel states [Nat. Commun. 11, 5613 (2020), Nat. Commun. 11, 4415 (2020)], and their study in FeGe is definitely a promising direction to pursuit. In this work, to further understand the edge state, we provide additional evidence in Fig. 3 showing that the edge state disappears when we raise the temperature above $T_{CDW}$ ~100K, further confirming the close relation between CDW and edge state as the main claim in this work.

We have checked half-step edges, including kagome lattice $Fe_3Ge$ and honeycomb lattice $Ge_2$ cases, and find that only $Fe_3Ge$ terminated half-step still has pronounced edge states.

**Question 5: It seems that the 2x2 charge order is only observed in conductance maps and not in topographic images - is this so? At what setpoint were the topographic images in Figs. 1d and 1e acquired? Do the authors see CDW contrast in topographic images at some bias?**

Answer: The 2×2 charge order in the dI/dV map is clear by eye. While the 2×2 charge order is not often clear in the topographic image by eye. A similar situation may be encountered for the $1 \times 4$ charge order in cuprates in several STM studies. We speculate that this is due to some short-range nature of the CDW coherence, but we agree that we do not fully understand the exact reason. We have stated the topographic image setup: "Topographic images in this work are taken with the tunneling junction set-up V = 60mV and I = 0.05nA."

We do detect slightly clearer 2×2 modulations in topographic images for certain bias, for instance, at -30meV:

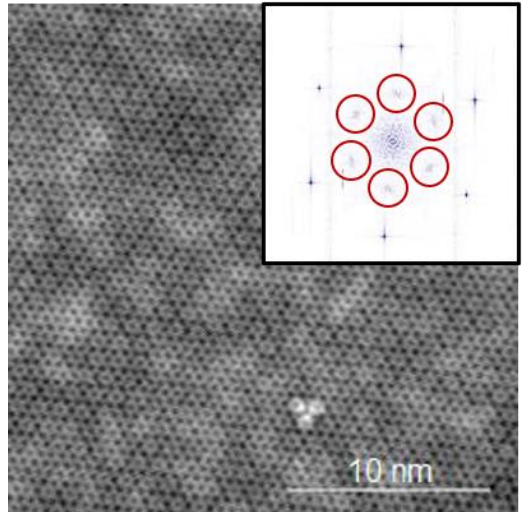

The corresponding Fourier transform in the inset shows signatures of 2×2 charge order. However, this is still not as clear as that presented in Fig. 1 using the dI/dV map data.

**Question 6: The 2x2 reconstruction is strong in Fig. 1f, which shows the spatial modulation of the local density of states at zero bias. If the 50meV structure measured by spectroscopy is actually the CDW gap, then we would not expect any (or at best a very weak) charge modulation amplitude at zero bias, which corresponds to the center of the CDW gap. Are there stronger charge modulations at energies between 0 and +/- 20 meV?**

Answer: In $AV_3Sb_5$ literature Refs. 7,18-26, for instance, dI/dV map at zero energy does similarly show strong 2 × 2 modulations. However, dI/dV map may not come purely from charge, but may mix certain lattice modulations that lead to the observed feature. The modulation strength does tend to become stronger between 0 and +-/20meV, as shown in Fig. 1f main panel, which is reproduced below:

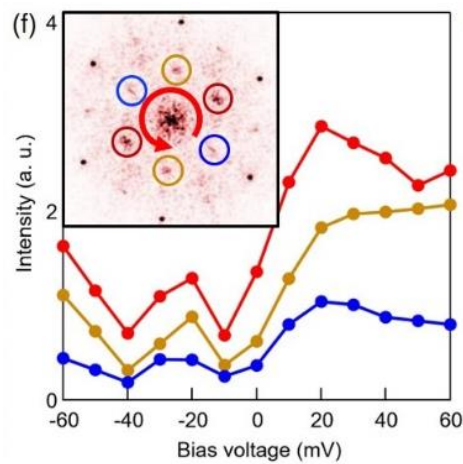

Figure captions: The main panel displays the modulation strength of the 2×2 charge order vector peaks from the Fourier transform of the dI/dV maps taken at different energies.

**Question 7: The tunneling conductance at $E_F$ at the step edge in Fig. 3c is more than 2 times larger than on the kagome lattice, giving rise to a large contrast at the step edge in Fig. 3a. While it is very difficult to judge,** **since all scales for tunneling conductance data are given in arbitrary units, we have the impression that the contrast in conductance maps between the step edge and the kagome lattice in spin polarized conductance maps (Figs. 2b and 2c) is lower than with a non-magnetic tip, when tip spin polarization is opposed to the kagome one and that the edge state is almost vanishing when tip and kagome spin polarizations are parallel. Did the authors attempt a comparison?**

Answer: The difference in edge state intensity in a spin-polarized study firstly indicates 1) the robustness of the edge state, 2) the edge state can be spin-polarized as well, which is not surprising for a magnetic system.

Initially, we were similarly puzzled by the observation that the polarization of the edge state seems to be opposite to the polarization of the kagome lattice. A crucial point to resolving this puzzle is: the spin-polarized map is taken at $E_F$ within the CDW gap, which makes it different from a conventional spin-polarized signal.

Suppose we have two kinds of electrons with spin-up and spin-down polarizations, while the CDW gap opens in the spin-up channel but not the spin-down channel. For spin-polarized tunneling, when we sense higher intensity at $E_F$, it indicates tunneling more to the spin-down channel, rather than the spin-up channel. If this is the case in our experiment, then it leads to the opposite spin-polarized tunneling behavior between the kagome lattice state and its edge state at $E_F$.

However, this is only one of the possible reasons, so we do not discuss this detail in the manuscript substantially. More broadly, the edge state is supposed to be chiral, and the chirality of the edge state should be reversed between adjacent kagome layers. We think it is an exciting topic that is worth substantial effort to demonstrate the spin-polarized and chiral nature of the edge state.